\documentclass[letterpaper,10pt]{article} 

\usepackage{opticameet3} %

\newcommand\authormark[1]{\textsuperscript{#1}}

\usepackage{amsmath,amssymb}
\usepackage[breaklinks,colorlinks=true,bookmarks=false,citecolor=blue,urlcolor=blue]{hyperref} %
\usepackage{pgfplots}
\pgfplotsset{compat=1.18}
\usepgfplotslibrary{colormaps}
\usetikzlibrary{spy}
\usetikzlibrary{external}
\usetikzlibrary{shapes.geometric}
\usepackage{subcaption}

\usepackage{comment}
\usepackage{cite}
\usepackage{siunitx}
\usepackage{wrapfig}
\usepackage{mathtools}
\usepackage[cal=cm]{mathalfa}

\definecolor{new_blue}{RGB}{93,169,233}
\definecolor{new_green}{HTML}{0FA3B1}
\definecolor{new_red}{HTML}{DD2D4A}
\definecolor{new_braun}{HTML}{79745C}
\definecolor{new_cherry}{HTML}{880D1E}
\definecolor{new_fildisi}{HTML}{F5EFED}
\definecolor{new_turqouise}{HTML}{508991}
\definecolor{new_purple}{HTML}{473198}
\definecolor{new_gray}{HTML}{67597A}

\definecolor{new_purple_soft}{HTML}{ACA2D2}
\definecolor{new_gray_soft}{HTML}{BCB5C4}
\definecolor{new_red_soft}{HTML}{DE828E}

\begin{document}

\title{Iteration-Dependent Scaled Min-Sum Decoding for Low-Complexity Key Reconciliation in CV-QKD}

\vspace{-5pt}
\author{Erdem Eray {C}il\authormark{*} and Laurent Schmalen}

\address{Communications Engineering Lab, Karlsruhe Institute of Technology, Karlsruhe, Germany\\}

\email{\authormark{*}\texttt{erdem.cil@kit.edu}} %

\setlength{\abovedisplayskip}{0pt}
\setlength{\belowdisplayskip}{0pt}

\vspace{-10pt}

\begin{abstract}
\vspace{0pt}
We introduce an iteration-dependent scaled min-sum decoding for low-rate LDPC codes in CV-QKD, achieving near-sum product algorithm performance with reduced complexity, and facilitating CV-QKD hardware implementation.
\end{abstract}

\vspace{-4pt}
\section{Introduction}
\vspace{-5pt}
Quantum computing has advanced remarkably in recent years, posing a significant threat to the current data encryption systems that rely on the computational hardness of the factoring problem \cite{Djordjevic_book}. Quantum key distribution (QKD)  is a physical-layer security scheme that provides secure keys to be used with a symmetric encryption scheme. Continuous-variable quantum key distribution (CV-QKD) is a promising technique for secure communication over long distances, as demonstrated by the experimental achievement of a cryptographic key exchange over more than 200 km \cite{208kmCV-QKD}.

However, implementing CV-QKD in hardware remains a challenge, especially in the information reconciliation step, where the two parties need to generate a common raw key from their measurements. This step typically involves the use of multi-edge type (MET) low-density parity-check (LDPC) codes \cite{9562244, Leverrier_CV_QKD_gaussian_mod, Mani_2021}, which require large block lengths and high decoding iterations to achieve satisfactory performance \cite{9562244}. These requirements result in a low decoder throughput, which reduces the secret key rate (SKR) compared to the achievable theoretical value. Therefore, low-complexity decoding algorithms are needed to optimize the system performance.

The scaled min-sum algorithm (MSA) is a frequently-used low-complexity decoding scheme, which approximates the sum-product algorithm (SPA) under the assumption of high signal reliability. This assumption holds for high-capacity channels, like optical channel, thus the performance gap between the MSA and the SPA is minimal. However, in the case of low-capacity channels such as quantum channels, this assumption is not met, leading to substantial deterioration in performance. Consequently, there is a need for a new low-complexity scheme tailored to CV-QKD operation.

In this work, we present a new method to compute the scaling coefficients for the scaled MSA that achieves near-SPA performance with lower complexity. We do this by rewriting the SPA check node (CN) update equation as a scaled MSA CN update equation and then estimating the scaling coefficients for each edge type and iteration. We evaluate the performance of our proposed iteration-dependent (ID) scaled MSA (ID-MSA) algorithm by decoding TBP LDPC codes with rates of $R=0.01$ and $R=0.1$ \cite{9562244}, which are suitable to be used in long-distance CV-QKD systems.
\vspace{-10pt}
\section{Check Node Update Approximation}
\vspace{-5pt}
In this section, we derive an approximation for the SPA CN update equation, which computes the output log-likelihood ratio (LLR) for each connected edge based on input LLRs received from its connected variable nodes (VNs).

The box-plus operator for input LLRs $L_1=\alpha_1\beta_1$ and $L_2=\alpha_2\beta_2$, where $\alpha_i$ and $\beta_i$ represent the sign and the magnitude of the LLR respectively, can be expressed as \cite[Ch. 5]{Ryan2009}
\begin{equation}
    L_1 \boxplus L_2 = \alpha_1\alpha_2 \Bigg( \min(\beta_1,\beta_2) + \underbrace{\log{\left(\frac{1+\exp{\left(-\big\lvert\beta_1+\beta_2\big\lvert\right)}}{1+\exp{\left(-\big\lvert\beta_1-\beta_2\big\lvert\right)}}\right)}}_{\coloneqq \mathrm{s}(\beta_1,\beta_2)} \Bigg) . \label{eqn:boxplus_section}
\end{equation}

To achieve the SPA performance with the MSA, we focus on the second term in (\ref{eqn:boxplus_section}), which represents the correction term of SPA over MSA. We denote this correction term as $\mathrm{s}(\beta_1,\beta_2)$. Without loss of generality, we assume that $\beta_1 \leq\beta_2$ and $\beta_2 = {a}\beta_1$ for any  $1 \leq {a} \in \mathbb{R}$. Under this assumption, the correction term $\mathrm{s}(\beta_1,\beta_2)$ can be reformulated as follows: 

\begin{eqnarray}
    \mathrm{s}(\beta_1,\beta_2) &=& \log{(1+\exp{(-({a}+1)\beta_1)})}- \log{(1+\exp{(-({a}-1)\beta_1)})} \nonumber\\
    &\overset{(i)}{=}&\sum_{k=1}^{\infty} \frac{(-1)^{k-1}}{k}\exp{(-k{a}\beta_1)}\left(\exp{(-k\beta_1)}-\exp{(k\beta_1)}\right) \nonumber\\
    &\overset{(ii)}{\approx}& 2 \sum_{k=1}^{\infty} (-1)^{k}\exp{(-k{a}\beta_1)}\beta_1 
    \overset{(iii)}{=} -\frac{2\ \beta_1}{1+\exp{({a}\beta_1)}} = -\frac{2\ \beta_1}{1+\exp{(\beta_2)}}.\label{eqn:correction_term}
\end{eqnarray}
Here, step $(i)$ involves expressing the logarithmic terms using the Taylor series. In step $(ii)$, by assuming small values of $k\beta_1$, we employ $\exp(x)\approx 1+x$. This assumption is reasonable since the density of the degree-1 VNs in typical low-rate codes used in CV-QKD is high. To give an example, in the TBP-LDPC code of rate 0.01 \cite{9562244}, 98.9\% of the CNs are connected to degree-1 VNs. It is important to note that this approximation holds for small values of $k\beta_1$, and the presence of $\exp{(-k\mathrm{a}\beta_1)}$ aids in reducing the error between the actual function and its approximation for large values. In step $(iii)$, we employ the  geometric series $1/(1+x)=\sum_{k=0}^\infty (-1)^kx^k$.

Replacing the correction term in (\ref{eqn:boxplus_section}) with the approximate correction term in (\ref{eqn:correction_term}) yields 
\begin{eqnarray}
    L_1 \boxplus L_2 &\approx& \alpha_1\alpha_2 \left( \min(\beta_1,\beta_2) - \frac{2\ \min(\beta_1,\beta_2)}{1+\exp{(\max(\beta_1,\beta_2))}} \right)
    = \alpha_1\alpha_2 \min(\beta_1,\beta_2) \tanh\left(\frac{\max(\beta_1,\beta_2)}{2}\right). \label{eqn:approx_box_plus}
\end{eqnarray}
Hence, the output LLR of the $i$th edge of CN $j$ with VN connections $\mathcal{M}(j)$  can be written as:
\begin{equation}
    L_i \approx \left( \prod_{k\in \mathcal{M}(j)/ \{i \}} \alpha_k \right) \beta_m \left(\prod_{\ell\in \mathcal{M}(j)/ \{i ,m\}} \tanh\left(\frac{\beta_\ell}{2}\right) \right) \label{eqn:app_CN_update},
\end{equation}
where $m$ is the index of the minimum $\beta$. While we derived (\ref{eqn:app_CN_update}) for low-rate codes, the same expression can also be used for high rate codes with a small performance penalty \cite{maganaDifferentPerspectiveApproach2012}. 

\vspace{-5pt}
\section{Iteration-dependent Scaling for the MSA}
In the previous section, we demonstrated that the SPA CN update equation can be effectively approximated using a scaled MSA CN update equation. The scaling factor, in this case, is determined by the product of hyperbolic tangent functions applied to the LLRs. Nonetheless, the computational demands of calculating this factor in real-time, involving hyperbolic tangent evaluations and multiplications, remain quite substantial. To mitigate this computational complexity, we introduce an alternative method: we propose using the expected value $\mathbb{E}\{\prod_\ell \tanh{ ( \beta_\ell/2 ) } \}$ for the MSA's scaling factor. This approach simplifies the method in \cite{zhouDEaidedANMSAEdge2021} to calculate the scaling coefficients that depend on both decoding iterations and edge type.

In the context of a long-distance CV-QKD system, it becomes evident that fixed scaling coefficients as in \cite{zhouDEaidedANMSAEdge2021}, which do not consider the input LLR values, are not sufficient. This observation is corroborated by (\ref{eqn:app_CN_update}), where the second-smallest LLR value dominates the scaling term. Additionally, transmission over the quantum channel leads to LLR distributions with low mean and high variance. Consequently, it is imperative to consider the second-smallest LLR value in order to improve the scaling coefficients and approach performance levels close to the SPA. This approach enables us to express the CN update equation of the ID-MSA for the $t$th iteration as follows:
\begin{eqnarray}
  L^{t}_i = \left( \prod_{k\in \mathcal{M}(j)/ \{i \}} \textrm{sign}(L_k) \right) \beta_m \mathrm{c}_i^t\left(\beta_{m'}\right), \label{eqn:main_eqn}
\end{eqnarray}
where $m'$ is the index of the second smallest $\beta$. To obtain the ID scaling term $\mathrm{c}_i^t\left(\beta_{m'}\right)$, we take the conditional expectation of the scaling factor in (\ref{eqn:app_CN_update}) given the second smallest $\beta$ value and assume independent LLRs. The ID scaling term is then given by
\begin{eqnarray}
  \mathrm{c}_{i}^t\left(\beta_{m'}\right) = \mathbb{E} \left\{ \prod_{\ell\in \mathcal{M}(j)/ \{i ,m\}} \tanh\left(\frac{\beta_\ell}{2}\right) \bigg\lvert \beta_{m'} \right\}
  = \tanh\left(\frac{\beta_{m'}}{2}\right) \prod_{\ell\in \mathcal{M}(j)/ \{i ,m, m'\}} \int_{x\in \mathbb{S}} \tanh\left(\frac{|x|}{2}\right) f_\ell^t(x) \mathrm{d}x, \label{eqn:coeff_calculation}
\end{eqnarray}
where  $\mathbb{S}=\{(-\infty,-\beta_{m'}] \cup [\beta_{m'},\infty)\}$ is the set of values that have a larger magnitude than the second minimum $\beta_{m'}$ and $f_\ell^t(x)$ is the probability density function of the LLR at iteration $t$ for the edge type $\ell$.   

\vspace{-5pt}
\section{Results and Evaluation}
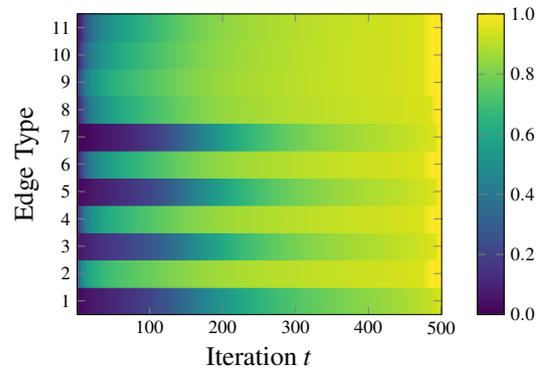
\begin{wrapfigure}{o}{0.45\textwidth}
    \vspace{-30pt}
    \tikzsetnextfilename{output-figure0}
\begin{tikzpicture}[scale=0.7]
        \begin{axis}[
            xlabel=\Large Iteration $t$,
            ylabel=\Large Edge Type ,
            colorbar,
            colorbar style={
                yticklabel style={
                    /pgf/number format/.cd,
                    fixed,
                    precision=1,
                    fixed zerofill,
                },
            },
            colormap/viridis,
            ytick={1,...,11},
            enlargelimits=false,
            axis on top,
            point meta min=0,
            point meta max=1,
            ]
            \addplot [matrix plot*,point meta=explicit] file [] {data/heatmap_data.dat};
        \end{axis}
\end{tikzpicture}
\captionsetup{font=small}
    \vspace{-5pt}
    \caption{Average values of $\mathrm{c}_i^t\left(\beta_{m'}\right)$ for $R=0.01$ TBP-LDPC code protograph \cite{9562244} }
    \label{fig:heatmap}
    \vspace{-20pt}
\end{wrapfigure}
In this section, we evaluate the performance of the ID-MSA to decode TBP-LDPC codes of rates $R=0.01$ and $R=0.1$ of block length $N$, as defined in \cite{9562244}. All algorithms perform 500 decoding iterations on the output of the binary-input additive white Gaussian noise channel with noise power $\sigma_n^2=N_0/2$.

To obtain the required statistics used in (\ref{eqn:coeff_calculation}), we perform an extrinsic information transfer chart analysis using the code protograph\cite{PEXIT}. Subsequently, we calculate look-up tables (LUTs) for $\mathrm{c}_i^t(\beta_{m'})$ for $t \in \{1,\dots,500\}$, and $i \in \{1,\dots,e\}$, where $e$ is the number of different edge types in the code graph and equals 11 and 8 for the rates 0.01 and 0.1, respectively, and $\beta_{m'}$ is quantized to 32 values. Figure \ref{fig:heatmap} shows how the mean scaling factor $\mathbb{E}_{\beta_{m'}} \{ c_i^t( \beta_m') \}$ changes with the decoding iteration for each edge type for the code of rate $R=0.01$. This figure clearly demonstrates the necessity of using multiple scaling factors in the MSA. To reduce the size and redundancy of the LUTs, we apply the K-means algorithm to obtain the final compressed LUT with 2600 entries for the decoding process.
\begin{figure*}[t!]
\centering
\begin{subfigure}{0.5\textwidth}
    \tikzsetnextfilename{output-figure1}
\begin{tikzpicture}[every pin/.style={fill=white},scale=0.8]
  \begin{axis}[
    grid=major,
    xlabel={ \text{$E_\text{s}/N_0$} (dB)},
    ylabel={Frame Error Rate},
    font = \large,
    ymode=log,
    legend pos=south west,
    grid=major,
    ymax=1,
    ymin=0.01,
    legend style={align=left},
    legend cell align=left, 
    xmin=-21.55,
    xmax=-15,
    axis line style =very thick,
  ]
    \addplot[blue,mark=o, line width=1.2] table [x=SNR,y=FER] {data/R_0p01_SP_double_precision.dat};
\addlegendentry{ \small SPA} %
\addplot[new_red,mark=square, line width=1.2] table [x=SNR,y=FER] {data/R_0p01_MS_double_precision.dat};
\addlegendentry{ \small MSA} %
\addplot[new_green,mark=x, line width=1.2] table [x=SNR,y=FER] {data/R_0p01_ID_MS_LUT_32_Kmeans_50_idx_Kmeans_500.dat};
\addlegendentry{ \small ID-MSA} %

    \coordinate (spypoint) at (axis cs:-21.7,0.099);
  \end{axis}
  
  \node [pin={[pin distance=0.5cm, pin edge={line width=0.5, black, dashed}]358:{
        {\begin{tikzpicture}[baseline,
          trim axis left,
          trim axis right,
        ]
          \begin{axis}[
            grid=major,
            xmin=-21.5,
            xmax=-21.15,
            ymin=0.05,
            ymax=1,
            ymode=log,
            width=3.6cm,
            height=3.6cm,
            tick label style={font=\tiny, xshift=0ex},
            axis line style = thick,
          ]
    \addplot[blue,mark=o, line width=1.2] table [x=SNR,y=FER] {data/R_0p01_SP_double_precision.dat};
    \addplot[new_green,mark=x, line width=1.2] table [x=SNR,y=FER] {data/R_0p01_ID_MS_LUT_32_Kmeans_50_idx_Kmeans_500.dat};
          \end{axis}
        \end{tikzpicture}}
      }
    },%
    draw, dashed, line width=0.5, minimum width=0.3cm, minimum height=2.93cm] at
   (spypoint) {};
\end{tikzpicture}
\caption{$R=0.01$ TBP-LDPC code \cite{9562244} with $N=998400$}
\label{fig:fer_snr_a}
\end{subfigure}%
\begin{subfigure}{0.5\textwidth}
 \tikzsetnextfilename{output-figure2}
\begin{tikzpicture}[every pin/.style={fill=white},scale=0.8]%
\begin{axis}[
    xlabel={ \text{$E_\text{s}/N_0$} (dB)},
    ylabel={Frame Error Rate},
    font = \large,
    ymode=log,
    legend pos=south west,
    grid=major,
    ymax=1,
    ymin=0.01,
    legend style={align=left},
    xmin=-11.4,
    xmax=-5,
    legend cell align=left, 
    axis line style =very thick,
]
\addplot[blue,mark=o, line width=1.2] table [x=SNR,y=FER] {data/R_0p1_SP_double_precision.dat};
\addlegendentry{ \small SPA} %
\addplot[new_red,mark=square, line width=1.2] table [x=SNR,y=FER] {data/R_0p1_MS_double_precision.dat};
\addlegendentry{ \small MSA} %
\addplot[new_green,mark=x, line width=1.2] table [x=SNR,y=FER] {data/R_0p1_ID_MS_LUT_32_Kmeans_50_idx_Kmeans_500.dat};
\addlegendentry{ \small  ID-MSA} %
   \coordinate (spypoint) at (axis cs:-11.53,0.099);
  \end{axis}
  \node [pin={[pin distance=0.45cm, pin edge={line width=0.5, black, dashed}]358:{
        {\begin{tikzpicture}[baseline,
          trim axis left,
          trim axis right,
        ]
          \begin{axis}[
            grid=major,
            xmin=-11.3,
            xmax=-10.9,
            ymin=0.01,
            ymax=1,
            ymode=log,
            width=3.6cm,
            height=3.6cm,
            tick label style={font=\tiny, xshift=0ex}, 
            axis line style = thick,
          ]
    \addplot[blue,mark=o, line width=1.2] table [x=SNR,y=FER] {data/R_0p1_SP_double_precision.dat};
    \addplot[new_green,mark=x, line width=1.2] table [x=SNR,y=FER] {data/R_0p1_ID_MS_LUT_32_Kmeans_50_idx_Kmeans_500.dat};
          \end{axis}
        \end{tikzpicture}}
      }
    },%
    draw, dashed, line width=0.5, minimum width=0.35cm, minimum height=2.93cm] at
   (spypoint) {};
\end{tikzpicture} 
\caption{$R=0.1$ TBP-LDPC code \cite{9562244} with $N=128000$}
\label{fig:fer_snr_b}
\end{subfigure}
\captionsetup{font=small}
\vspace{-15pt}
\caption{Performance comparison of sum-product algorithm (SPA), scaled min-sum algorithm with the factor $0.75$ (MSA), and the proposed iteration-dependent MSA (ID-MSA)}
\label{fig:fer_snr}
\vspace{-25pt}
\end{figure*}
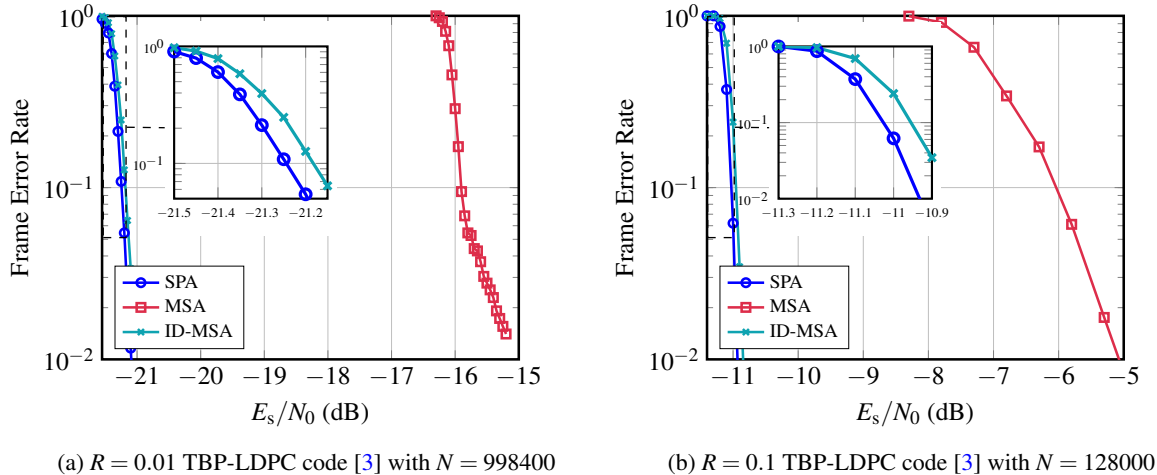

We compare the performance of the algorithms at an FER of 0.1, since the secret key rate of the system is typically maximized at this value \cite{208kmCV-QKD}. Fig. \ref{fig:fer_snr_a} shows that the SP algorithm outperforms the scaled MS algorithm by $5.2 \ \si{dB}$ for the $R=0.01$ code. This large gap reduces the SKR significantly, making the MSA unsuitable for CV-QKD applications. However, by using iteration-dependent scaling coefficients, we can reduce this gap to only $0.059 \ \si{dB}$, achieving a performance close to that of the SP algorithm. Similarly, Fig. \ref{fig:fer_snr_b} shows that the ID-MSA also performs close to the SP algorithm for the $R=0.1$ code, with a gap of only $0.068 \ \si{dB}$. Thus, the ID-MSA is a promising candidate for low-complexity key reconciliation in long-distance CV-QKD scenarios.

To compare the decoding complexity of ID-MSA with existing algorithms,  we consider the CN update as described in \cite{LambdaMin}, which employs (\ref{eqn:boxplus_section}) for the 3 LLRs with the smallest magnitudes to approach SPA performance. This algorithm requires two evaluations of (\ref{eqn:boxplus_section}) causing 2 table lookups for each CN update. In contrast, our proposed ID-MS algorithm requires only one lookup. Hence the decoding complexity of the proposed algorithm is less than the current approaches that can achieve near-SPA performance. 

\vspace{-5pt}

\section{Conclusion}
The ID-MSA proposed in this paper exhibits more than $5 \ \si{dB}$ performance improvement over the MSA, achieving near-SPA decoding performance with reduced computational complexity. Simulation results validate the efficacy of our approach and demonstrate its effectiveness across a wide range of low rates. This makes it a promising solution for efficient low-rate decoding, particularly suitable for CV-QKD systems and other related applications.

\vspace{-5pt}
\footnotesize{
\section*{Acknowledgement}
This work was funded by the German Federal Ministry of Education and Research (BMBF) under grant agreement 16KISQ056 (DE-QOR).

}

\vspace{-5pt}
\bibliographystyle{opticajnl}
\bibliography{sample}

\end{document}